\documentclass[aps,%draft,
superscriptaddress,amsmath,amssymb,showpacs,twocolumn]{revtex4}

\usepackage{graphicx}
\usepackage{epstopdf}
\usepackage{hyperref}
\usepackage{epsfig}
\usepackage{color}
\usepackage{psfrag}
\usepackage{float}
\usepackage{bbold}
\usepackage{tikz}
\usepackage[caption=false]{subfig}

\newcommand{\rem}[1]{}

\newcommand{\refE}[1]{Eq.~(\ref{#1})}
\newcommand{\refF}[1]{Fig.~\ref{#1}}
\begin{document}

\title{Scaling laws for the bifurcation-escape rate in a nanomechanical resonator}

\author{M. Defoort}

\affiliation{
Universit\'e Grenoble Alpes, CNRS Institut N\'EEL, \\
 BP 166, 38042 Grenoble Cedex 9, France}

\author{V. Puller}

\affiliation{
Univ.~Bordeaux, LOMA, UMR 5798, F-33400 Talence, France \\
CNRS, LOMA, UMR 5798, F-33400 Talence, France}

\author{O. Bourgeois}
\affiliation{
Universit\'e Grenoble Alpes, CNRS Institut N\'EEL, \\
 BP 166, 38042 Grenoble Cedex 9, France}

\author{F. Pistolesi}

\affiliation{
Univ.~Bordeaux, LOMA, UMR 5798, F-33400 Talence, France \\
CNRS, LOMA, UMR 5798, F-33400 Talence, France}

\author{E. Collin}
\affiliation{
Universit\'e Grenoble Alpes, CNRS Institut N\'EEL, \\
 BP 166, 38042 Grenoble Cedex 9, France}

\begin{abstract}
We report on experimental and theoretical studies of the fluctuation-induced escape time from a metastable 
state of a nanomechanical Duffing resonator in cryogenic environment. 
By tuning in situ the non-linear coefficient $\gamma$ 
%instead of the driving force 
we could explore a wide range of 
the parameter space around the bifurcation point, where the metastable state becomes unstable. 
We measured in a relaxation process the distribution of the escape times.
We have been able to verify its exponential distribution and extract the escape rate $\Gamma$. 
We investigated the scaling of $\Gamma$ with respect to the distance to the bifurcation point and $\gamma$, finding an unprecedented quantitative agreement with the theoretical description of the stochastic problem. 
Simple power scaling laws turn out to hold in a large region of the parameter's space, as anticipated by recent theoretical predictions.
%
%For both control parameters, the analytic power scaling laws turn out to hold in the whole parameter space explored, even far from the bifurcation point. 
%NEW:
These unique findings, implemented in a model dynamical system, are relevant to all systems experiencing under-damped saddle-node bifurcation.
\end{abstract}


\newcommand{\Teff}{D} %{T_{\rm eff}}

\maketitle

Transition from a metastable to a stable state is a phenomenon of ubiquitous interest in science: 
in thermal equilibrium it is the essence of the activation law in chemistry \cite{arrhenius,kramers_brownian_1940}, it underlies nucleation in phase transitions, magnetization reversal in molecular magnets \cite{novak_magnetic_1995}, biological switches in cells behavior \cite{ozbudak_multistability_2004} or RNA dynamics \cite{RNA_2004}, transitions of Josephson junctions \cite{turlot_escape_1989} or fluctuations in SQUIDs \cite{kurkijarvi_intrinsic_1972}, the list being obviously non-exhaustive.
More recently the study of escape statistics has been possible also for out-of-equilibrium {\em dynamical} systems like Penning traps \cite{lapidus_stochastic_1999}, Josephson junctions \cite{siddiqi_rf-driven_2004}, and nano-electromechanical 
systems \cite{aldridge_noise-enabled_2005,stambaugh_noise-activated_2006, chan_activation_2007,chan_paths_2008, unterreithmeier_nonlinear_2010}: the state-switching effect is extensively used in bifurcation amplifiers, with for instance state-of-the-art quantum bit readout schemes \cite{boulant_quantum_2007}.
In most of these cases the escape time distribution is exponential and the rate $\Gamma$ characterizes completely the phenomenon. 
Analytical solutions \cite{hanggi_reaction-rate_1990} of the dynamical equations show that its value depends exponentially on a parameter $\Teff^{-1}$, that coincides with the (inverse of the) temperature for equilibrium systems and more generally is related to the power spectrum of the relevant fluctuations.
One can then write: 
\begin{equation}
	\Gamma = \Gamma_0 \exp ( -E_a / \Teff ) ,
	\label{mainEq}
\end{equation}
where the prefactor $\Gamma_0$ is assumed to depend very weakly 
on $\Teff$, and $E_a$ in analogy with a potential system can be called activation energy: it parametrizes the distance to the unstable point. 
For out-of-equilibrium systems a central theoretical result is the paper by Dykman and Krivoglaz \cite{dykman_theory_1979},
that found an explicit expression for $E_a$ and $\Gamma_0$ for a generic dynamical system close to 
the bifurcation point, where the line of metastable states joins the line of unstable ones.
It predicts universal power laws dependence of $E_a$ and $\Gamma_0$ on the distance from the bifurcation point in terms of $|\omega-\omega_b | $, where 
$\omega$ is the driving frequency of the dynamical system and $\omega_b$ is its  
bifurcation value.

Direct experimental measurement of the escape time and study of the 
dependence of $E_a$ and $\Gamma_0$ over a wide range of 
%NEW:
a system's parameters
is not a trivial task, since the exponential dependence of the escape time 
makes it either too long or too short for a reasonable observation protocol.
For dynamical systems the resonating period fixes a lower bound on the time.
% NEW:
Nano-mechanical resonators with resonance frequency in the MHz range are thus 
a prominent choice to investigate the bifurcation instability of Duffing oscillators: 
they are high frequency dynamical systems with a high quality factor for which the distance to the bifurcation point can be directly controlled.

%NEW:
In the analysis of switching and reaction rates, three problems can thus be distinguished: obtaining the exponent $E_a$, the prefactor $\Gamma_0$, and their respective scalings for systems away from thermal equilibrium.
The exponent has been the first subject of interest, with the early work of Arrhenius \cite{arrhenius}. The prefactor has then been addressed by Kramers later on \cite{kramers_brownian_1940}, while finally the scaling of both for dynamical systems has been derived by Dykman \cite{dykman_theory_1979}.
It is actually in micro and nano-mechanical systems that a measurement of the power law dependence of $E_a$ with respect to the distance from the bifurcation point has been performed, giving the predicted value within experimental error \cite{aldridge_noise-enabled_2005,stambaugh_noise-activated_2006}.
Nevertheless, the activation energy has been claimed to match theory at best within a factor of 2 due to injected noise calibration \cite{aldridge_noise-enabled_2005}.
To our knowledge no attempts have been done to obtain a more quantitative verification
of the predictions of Dykman and Krivoglaz \cite{dykman_theory_1979}, in particular for the scaling law of the prefactor $\Gamma_0$ and the dependence to the Duffing non-linear coefficient $\gamma$ of both $\Gamma_0$ and $E_a$.
%
%NEW:
Answering the three above mentioned problems {\it together} is thus the aim of our work, using a unique nano-mechanical implementation of the bifurcation phenomenon.

In this Letter we report on experimental and theoretical investigations of the dependence 
of $E_a$ and $\Gamma_0$ on the system parameters for a driven nano-mechanical oscillator in the non-linear regime in presence of a controlled noise force. 
It is well known that for a sufficiently strong non-linear term the system admits for some values
of the driving frequency a metastable solution.
By measuring the escape rate for a wide range of parameters we could verify the validity 
of the power scaling laws predicted by Dykman and Krivoglaz for both $E_a$ and $\Gamma_0$.
Remarkably, we found that the scaling holds experimentally in a much larger region of the parameter 
space than the one for which the theory of Ref.~\cite{dykman_theory_1979} has been derived.
%
%edd's rewriting
%To our knwowledge, only two theoretical works addressed directly the question of the analytic calculation's validity range and power law prescriptions \cite{DykScaling,kogan2008}.
%Fabio's
Concerning the $E_a$ dependence on detuning, the possibility of an extended region of scaling was discussed in Refs. \cite{DykScaling,kogan2008}. 
Performing the full numerical simulation of the stochastic problem adapted to our device parameters we found that experiment and theory are in excellent quantitative agreement.
%, even far from the bifurcation point.
%
%We thus provide an unprecedented experimental verification of the recent theoretical predictions on under-damped saddle-node bifurcation \cite{DykScaling,kogan2008}.
%
%We thus provide a quantitative analysis of the escape process for the Duffing oscillator.
%
%This scaling result is specific to the Duffing oscillator, but the coincidence of the power law with the universal exponent may suggest that there is a more general origin applying to any saddle-node bifurcation system.

\vspace*{1mm}

% MAIN TEXT OF THE PAPER 

% Experimental device

%%%%%%%%%%%%%%%%%%%%%%%%%%%%%%%%%
%
%          Figure 1
%
%%%%%%%%%%%%%%%%%%%%%%%%%%%%%%%%%
%
%
\begin{figure}
\includegraphics[height=7.3 cm]{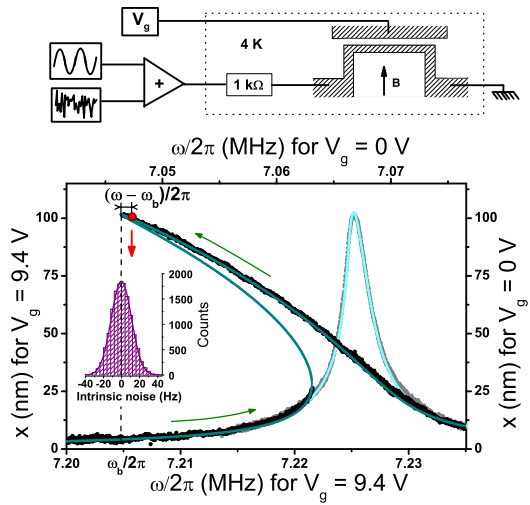}
\caption{
\label{hyst} 
(Color online) 
Top panel: Schematic of the experimental setup with the nano-resonator structure. Bottom panel: Linear and Duffing resonances (respectively grey and black points, with top-right and bottom-left axes).
The lines show the fit.
The nonlinear resonance is for $V_g$ = 9.4$~$V, which shifts the resonance frequency and opens an hysteresis (thin green arrows highlight upward and downward sweeps). 
The relaxations occur at a detuning $\omega-\omega_b$ from the bifurcation frequency (red point and vertical arrow). Inset: Gaussian distribution histogram of the measured intrinsic  frequency fluctuations.}
\end{figure}

\begin{figure}
\includegraphics[height=5.2 cm]{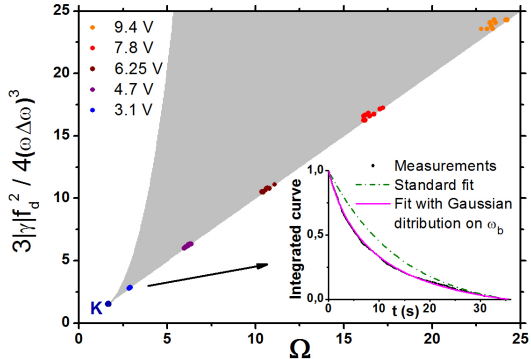}
\caption{\label{ParamSpace} (Color online) Bifurcation parameter space (normalized driving force versus $\Omega=2|\omega-\omega_0|/\Delta \omega$). The grey area is the NEMS bistability regime where the right edge is the transition from a high amplitude oscillation to a low one (the left edge is the opposite) and K is the spinode point where hysteresis starts to open. We show within the bistability the data points at different voltages $V_g$. Inset: typical low $V_g$ relaxation curve obtained with about 1000 relaxations, and fit with and without fluctuations on $\omega_b$.}
\end{figure}

The experiment is performed on a unique goalpost (depicted in top graph of \refF{hyst}) aluminum-coated silicon nano-electro-mechanical resonator. 
It consists in two cantilever feet of length 3 $\mu$m linked by a paddle of length 7 $\mu$m, 
all about 250 nm wide and 150 nm thick for a 
%
%The silicon thickness is 150 nm while the conducting aluminum layer is 30 nm, for a
 total mass 
$m =$ 1.25 $10^{-15}$ kg \cite{collin_-situ_2012}.
The experiment is performed at $4.2$ K in cryogenic vacuum (pressure$~< 10^{-6}$ mbar). 
The motion is actuated and detected by means of the magnetomotive scheme \cite{clelandmagneto}, 
with a magnetic field $B < 1$ T  
and a gate electrode is also capacitively coupled to the nanomechanical device (gap about 100 nm)
\cite{collin_-situ_2012}.
The resonator admits large distortions (in the hundred nm range) 
to be attained while remaining
intrinsically extremely linear \cite{collin_tunable_2011}, while a well-controlled non-linearity can be generated by means of a DC gate voltage bias $V_g$ \cite{kozinsky_tuning_2006}. 
This distinctive feature enables to tune the global non-linearity of our device without changing the displacement amplitude.
Using an adder we apply both a sinusoidal drive and a noise voltage from a voltage source generator.
The resulting electric signal together with a 1 kOhm bias resistor is used to inject an AC current through the goalpost and generates both driving and controllable (zero average) noise forces on the resonator.
More information on the calibration and experimental details can be found in Refs. \cite{collin_tunable_2011,collin_-situ_2012}.
The resulting equation of motion for the resonator displacement $x$ reads:
\begin{equation}
	\label{langevin}
	\ddot x+ \Delta \omega \dot x+ \omega_0^2 x + \gamma x^3 = f_d \cos(\omega t)+f_n(t)
\end{equation}
with $\omega_0/2 \pi$=7.07 MHz the resonance frequency, 
$\Delta \omega/2\pi$=1.84 kHz the linewidth, and $f_d$ and $f_n$ the drive and noise forces divided by the mass of the resonator.
We fix the drive force so that $mf_d~$=$~$65$~$pN, leading to a constant maximal displacement amplitude of 100$~$nm.
%reply referee 4
As can be deduced from our characterizations \cite{collin_-situ_2012}, this amplitude is small enough to guarantee that nonlinear damping mechanisms such as discussed in Refs. \cite{NonlinDamping,parametric} are small (see comment in the discussion section).
The noise force signal is filtered so that the force spectrum $\int dt e^{i \omega t}\langle f_n(t) f_n(0) \rangle_\omega=2D$ is constant over a bandwidth of 1 MHz around 7 MHz.
The Duffing coefficient $\gamma$ scales as $V_g^2$ and is for us negative \cite{collin_tunable_2011}.
At fixed driving force, the system admits two amplitudes of oscillation for sufficiently large $|\gamma|$ as shown in \refF{hyst} (bistability).
%, where a typical measured bistable resonance curve is displayed. 
%
By fitting with the standard Duffing expressions \cite{collin_addressing_2010} the parameters $\Delta \omega$, $\omega_0$ and 
$\gamma$ together with the bifurcation frequency $\omega_b$ can be obtained with a good accuracy.
The experiment is then performed by sweeping $\omega$ from the stable regime ($\omega>\omega_0$) down to the edge of the hysteresis at a given value of $\omega-\omega_b$ in the high amplitude state (see \refF{hyst}).
%reply referee 4
The sweeping rate (a few Hz/s) is an important parameter which should both guarantee adiabaticity of the sweep and high accuracy in the measurement \footnote{We can estimate adiabaticity using theoretical expressions from Ref. \cite{NonlinDamping}. Eqs. (50), (53) and (57) give the shift in bifurcation frequency $\delta_{err}$ induced by finite sweep rate. We obtain $\delta_{err} = 2 \pi \, 0.5~$Hz at most with our experimental parameters, which means that the error in the resonance position is less than 20$~$ppm of the Duffing frequency shift itself. At the same time, the critical slowing down time $\tau_{sd}$ can be estimated from Eq. (52). We obtain $\tau_{sd}$ smaller than 8$~$ms for all our settings, which shall be compared to the smallest recorded bifurcation time of order 40$~$ms.  }.
Finally, the escape time from the metastable state is detected when the measured displacement amplitude 
falls below an appropriate threshold value.
Typically  $10^3$ escape events are recorded for each set of parameters.
The experiment has been repeated for three different values of the noise forces $f_n$,  
three different detunings $\omega-\omega_b$ (up to 5$\%$ of the hysteresis), and 
five different values of $V_g$ (and  thus of $\gamma$), for a total of 45 escape histograms.
The resulting settings are summarized 
in \refF{ParamSpace}.
%

%%%%%%%%%%%%%%%%%%%%%%%%%%%%%%%%%
%
%          Figure 2
%
%%%%%%%%%%%%%%%%%%%%%%%%%%%%%%%%%
%

\vspace*{1mm}

\begin{figure}
\includegraphics[height=6.3 cm]{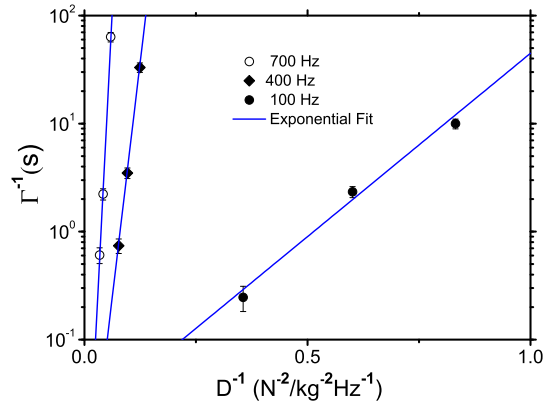}
\caption{\label{D} (Color online) Escape time as a function of $D^{-1}$ for $V_g=9.4~$V at different detunings $\omega-\omega_b$ from the bifurcation point.}
\end{figure}

For each data measurement, the experimental value of $\omega_b$ might slightly differ from the one 
obtained by the initial fit.
This problem is detected by sweeping relatively rapidly $\omega$ (tens of Hz/sec) through the bifurcation point 
and measuring the escape value $\omega_b$ prior to each relaxation-time acquisition.
A typical histogram of the distribution of $\omega_b$ is shown in the inset of \refF{hyst} for $V_g=9.4~$V. 
It has Gaussian form with a half-width $\sigma$ in the range of tens of Hz.
This tiny spread ($10^{-6}$ to $10^{-5}$ of $\omega_b$) is due to low-frequency intrinsic fluctuations of the resonating frequency, which actual origin is still under debate \cite{KingPRB2012, ZhangArXiv, LeeuwenArXiv}.
% \footnote{Nanomechanical devices are known to suffer from frequency noise. %, displaying even time to time characteristic telegraph noise. 
%The spectral density of our frequency noise appears to scale roughly as $1/f$.% between our two experimental cutoffs (limited at low frequency by the total acquisition time of about a day and at high frequency by the acquisiton time step of 40$~$ms).}.
%
Even if extremely small, due to the high sensitivity of the bifurcation phenomenon 
%on a shift of $\omega_b$, we have to take this effect into account in order to extract with high accuracy the escape rate for each set of data.
%
the fluctuations of $\omega_b$ modify slightly the value of $\Gamma$ at each measurement, and we have to take this effect into account.
The escape exponential distribution has thus to be averaged over these fluctuations. 
For  $|\omega-\omega_b | \gg \sigma$ one can expand this dependence:
 $\Gamma(\omega-\omega_b-\epsilon)=\Gamma+\Gamma' \epsilon+\dots$,
where $\epsilon$ is the Gaussian-distributed shift of $\omega_b$.
This gives the following distribution for the escape times:
\begin{equation}
P(t)=\Gamma e^{-\Gamma t} 
	\int {d \epsilon \over \sigma \sqrt{2 \pi} } e^{-\epsilon^2/(2\sigma^2)-\Gamma' \epsilon t}
	\, .
\end{equation}
Fitting it to the data with the method of Kolmogorov-Smirnov \cite{KolmoSmirn}, to avoid losses of information 
due to histogram binning, the two independent parameters of the distribution, $\Gamma$ 
and the product $\Gamma' \sigma$, can be obtained.
A typical curve is shown in the inset of \refF{ParamSpace}.
Note that this procedure does not need any hypothesis on the explicit functional dependence of $\Gamma$ on $\omega_b$.
On the other hand the procedure breaks down for too small detunings, and we thus need to drop 
the data for four values of the detuning.
We can then verify the validity of \refE{mainEq} for the system at hand by plotting  $\log \Gamma$ as 
a function of $1/D$ (see \refF{D}).
The linear fit gives $E_a$ and $\Gamma_0$.
The absolute experimental definition of the noise level is difficult, and we introduce a calibration factor $C$ (close to 1) between $D$ and the nominal injected noise power. Note that it simply amounts to multiply $E_a$ by $C$, thus leaving the scaling dependence unmodified. The value of $\Gamma_0$ is not affected by this calibration either.
%added for referee 4
%
More experimental details can be found in Ref. \cite{MartialPHD}.
%
%%%%%%%%%%%%%%%%%%%%%%%%%%%%%%%%%%%%%%%%%%%
%
%
%                 Fig 3
%
%
%%%%%%%%%%%%%%%%%%%%%%%%%%%%%%%%%%%%%%%%%%%

In order to extract the scaling dependence of $E_a$ and $\Gamma_0$ on the detuning and the non-linear parameter $\gamma$
it is convenient to recall the predictions that can be obtained following Ref.~\cite{dykman_theory_1979}. 
Let us rescale the detuning by defining  $\Omega=2|\omega-\omega_0|/\Delta \omega$ with 
$\Omega_b=2|\omega_b-\omega_0|/\Delta \omega$.
For $\Omega_b\gg \sqrt{3}$ (that holds for all the data of our experiment) one obtains that
$\Omega_b \approx 3 |\gamma| f_d^2/(4 \omega^2\Delta \omega^2)$ with 
the parameters in \refE{mainEq} reading 
\footnote{These expressions are obtained following the method of Ref. \cite{dykman_theory_1979}. Note that in our work the bifurcation is analysed as a function of frequency detuning and nonlinear parameter (not applied force).}:
\begin{equation}
	\label{dykmanFormula}
	E_a=\frac{2 f_d^2}{3 \Delta\omega}  {|\Omega-\Omega_b |^{3/2} \over \Omega_b^{5/2}},
	\,
	\Gamma_0=\frac{\Delta \omega}{2} \frac{|\Omega-\Omega_b |^{1/2} \Omega_b^{1/2}}{2\pi}.
\end{equation}
The basic assumptions to obtain these expressions are that $E_a/D \gg 1$ in order to keep the escape a rare event, and to be able to reduce this two-dimensional problem (amplitude and phase) into a one-dimensional one. 
This second condition (much less appreciated in the literature) is only verified when the driving frequency $\omega$ is in a tiny region close to the bifurcation point $\omega_b$ and far from the frequency for which the amplitude is maximum. 
In this region, one of the eigenvalues of the linearized dynamical equations of motion vanishes, which induces a slow motion in the direction of the relative eigenvector. 
On the other hand when $\omega$ is such that the amplitude is maximal, the two eigenvalues coincide, inducing fully two-dimensional fluctuations.
Thus beyond this point the approximation used to obtain \refE{dykmanFormula} breaks down.
This condition reads $4\Omega_b |\Omega-\Omega_b | \ll 1$.
%
%%%%%%%%%%%%%%%%%%%%%%%%%%%%%%%%%%%%%%%%%%%
%
%
%                 Fig 4
%
%
%%%%%%%%%%%%%%%%%%%%%%%%%%%%%%%%%%%%%%%%%%%
\begin{figure*}
\begin{subfloat}
  \centering
  \includegraphics[height=6.15 cm]{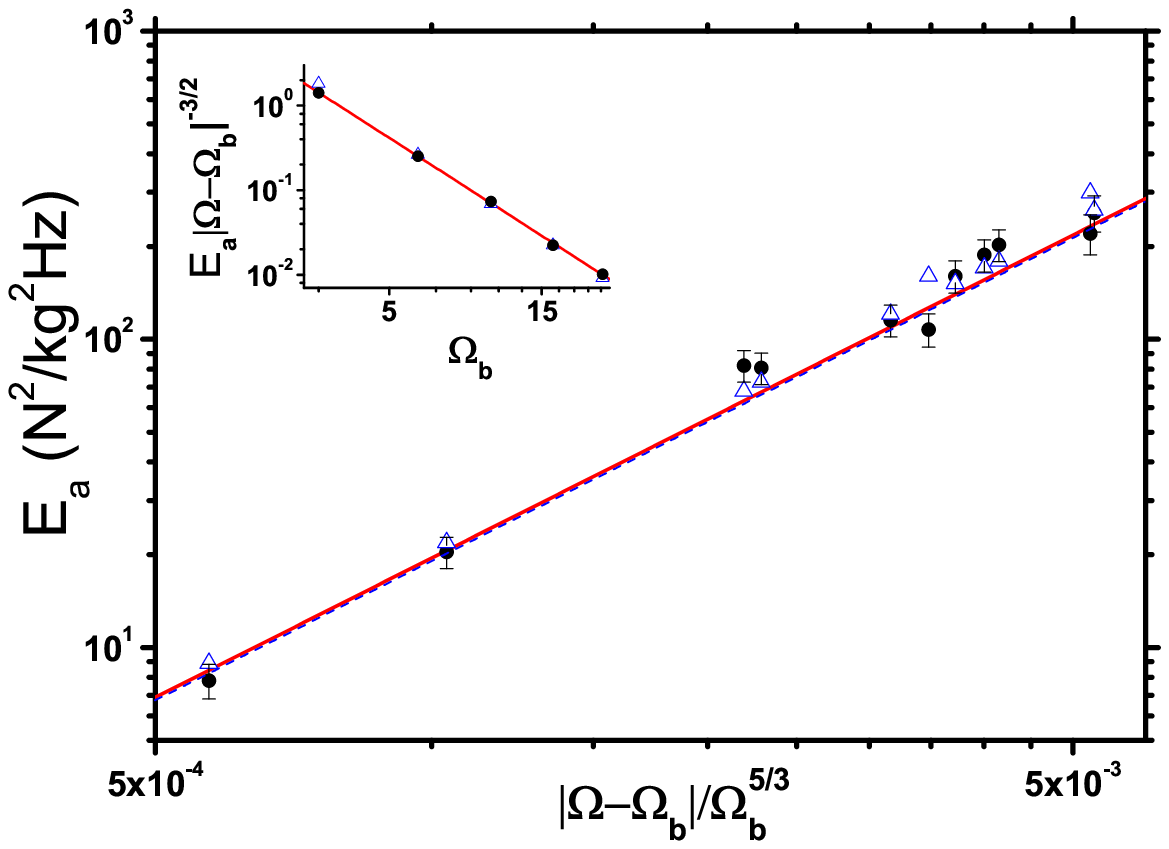}
  \label{Ea}
\end{subfloat}
\begin{subfloat}
  \centering
 \includegraphics[height=6.15 cm]{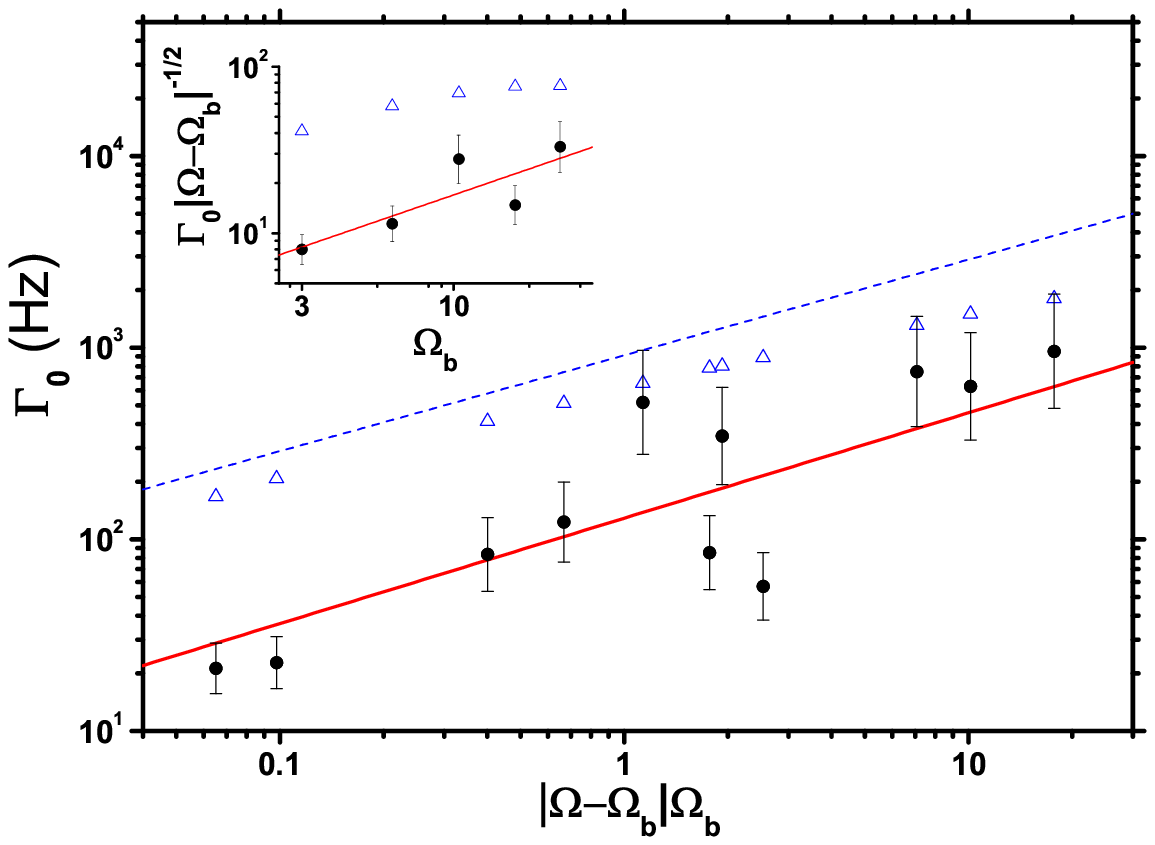}
  \label{Gamma}
\end{subfloat}
\caption{
(Color online) Scaling plots for  $E_a$ (left) and  $\Gamma_0$ (right) with respect to detuning.
The full circles indicate the experimental points, the open (blue) triangles the prediction of the full numerical 
simulation, the (red) full lines the linear fit to the data, and the dashed (blue) lines the prediction of \refE{dykmanFormula}.
Insets: scaling with the non-linear parameter $\Omega_b$. 
%added Eddy's
%Note that error bars on the left panel correspond to $\pm 20~\%$ on Log $\Gamma_0$.
}
\label{EG}
\end{figure*}
In the experiment we performed this quantity ranged uniformly between 0.13 to 71, thus a part of the data where well outside the range of the expected validity 
of \refE{dykmanFormula}, enabling to investigate the behavior of $\Gamma$ in a region where no present analytical prediction exists.
As explained, the expressions for $E_a$ and $\Gamma_0$ in \refE{dykmanFormula} depend only 
on the detuning and the non linear coefficient (through $\Omega_b$), 
the other parameters being the same for all data points.
To test the validity of Dykman-Krivoglaz expressions, we produce a scaling plot, where 
the logarithm of  $E_a$ and $\Gamma_0$ are plotted as a function of 
$|\Omega-\Omega_b|/\Omega_b^{5/3}$ and $|\Omega-\Omega_b|\Omega_b$ (see \refF{EG}).
A remarkable scaling is then observed in all the experimental range, with a fitted slope 
as a function of the detuning of $1.53 \pm 0.04$ and $0.55\pm0.2$, for $E_a$ and $\Gamma_0$ respectively. 
This matches the analytic predictions by Dykman and Krivoglaz, and we use this good agreement to define the noise source calibration factor $C$:
scaling $D$ by $C$ the prediction of \refE{dykmanFormula} coincides with the 
fitted value for $E_a$ (dashed line in \refF{EG} left panel).
%
%This is done to obtain a better estimate of $E_a$ in view of the comparison with the full
%numerical model in the following.  
%
The dependence on the non-linear parameter $\Omega_b$ could also be tested for both quantities. 
It is shown in the insets of \refF{EG} and gives fitted slopes of $-2.43\pm0.05$ and $0.6\pm 0.1$, again in excellent agreement with \refE{dykmanFormula}.
%
%

%
%As a matter of facts, we are not aware of previous attempts to compare experimental 
%measurement of the prefactor with existing theory.
\vspace*{1mm}

To better understand this remarkable %and unexpected 
scaling in such a large parameter region
we solved numerically the stochastic problem. 
This can be done by introducing the complex slow amplitude $z(t)$ defined as 
$x(t)=z(t)e^{i \omega t}+z(t)^* e^{-i\omega t}$ and then convert the Langevin \refE{langevin}
to a Fokker-Planck equation $\partial_\tau P= {\cal L} P$ for the probability density $P(u,v, \tau )$ of the real and imaginary part 
of $z=(3|\gamma|/\Delta \omega)^{1/2}(u+iv)$ as a function of the dimensionless time $\tau=t\Delta \omega$.  
The escape rate from a given domain can be calculated by solving the equation 
${\cal L}^\dag \tau(u,v)=-1$ with zero boundary condition at the border of the domain \cite{hanggi_reaction-rate_1990}.
This gives the average time needed to reach the border starting at $(u,v)$.
The equation reads explicitly:
\begin{equation}
	\label{firstPassage}
		[
		{\cal D} (\partial^2_u +\partial^2_v)
		-f_u \partial_u
		-f_v\partial_v] \tau=-1 
	\,,
\end{equation}
with ${\cal D}=3|\gamma| D/(8 \omega^3 \Delta \omega$), $f_u=u+v(u^2+v^2)-\Omega$, $f_v=v-u(u^2+v^2)-\Omega-F_d$, and 
$F_d=f_d (3|\gamma|)^{1/2}/[2(\omega \Delta \omega)^{3/2}$].
\refE{firstPassage} can be solved numerically \cite{pistolesi_self-consistent_2008} to obtain the average escape time that coincides with the inverse of the sought Poissonian rate. 
The numerical results for $E_a$ and $\Gamma_0$ are shown in \refF{EG} in open (blue) triangles.

One can see that the exact (numerical) result has the same power law dependence as the analytical results (dashed line), even where the approximate theory is not supposed to hold.
Quantitative agreement between experiment and theory on $E_a$ is obtained with $C \approx 1.3$, thus validating the experimental noise amplitude calibration to within 15$~$\% which is remarkable. Note that the simulation does not contain any other free parameter, which are all experimentally known to better than 5$~$\%.
Concerning $\Gamma_0$, we are not aware of previous attempts to compare 
this quantity to the theoretical predictions.
The agreement with the full theory is within a factor of about 3, which is remarkable given the logarithmic precision on this parameter.
%new
We speculate that these discrepancies could arise from the actual algorithm used to extract $\Gamma_0$, or from more fundamental reasons like extra (non-Duffing) nonlinearities appearing in \refE{langevin} (i.e. non-linear damping, or non-cubic restoring force terms) \footnote{We tried to quantify the impact of nonlinear damping and of gate coupling non-Duffing contributions on the dynamics equation.  Using a quadratic fit to describe nonlinear damping \cite{collin_addressing_2010, parametric}, one can estimate the $p$ parameter of Ref. \cite{NonlinDamping} to be at worst about $0.06$. Following the calculation procedure of Ref. \cite{dykman_theory_1979}, the alteration of the energy potential $E_a$ is then at worst about 20$~\%$. From the gate capacitance Taylor series coefficients of  Ref. \cite{collin_-situ_2012}, we calculate that the "effective" Duffing nonlinear parameter measured in a frequency-sweep experiment could be modified by about 4$~$\% with respect to the value computed from the actual $x^3$ restoring force term, which is small.}.
%The absolute discrepancy could arise from the actual algorithm used to extract $\Gamma_0$, or from more fundamental reasons like extra (non-Duffing) nonlinearities coming into play.
%We find a (logarithmic) agreement with the full numerical solution within a factor of 3.
%
%Fabio's
% do we really need to say that? If Gamma0 could be wrong, then Ea as well?
%The shift towards higher values may be due to the effect of an additional 
%small non-linearity that could systematically affect all the data, and in particular
%$\Gamma_0$ which is more difficult to extract. 

\vspace*{1mm}

In conclusion, we have investigated the escape dynamics close to the bifurcation 
point for a nanomechanical resonator in the Duffing non-linear regime measured 
at cryogenic temperatures.
Using a {\em single ideally tunable system}, we have: 
{\em (i)} Measured the escape rate $\Gamma$ as a function of the noise 
amplitude $D$, the detuning to the bifurcation point $\omega-\omega_b$, and the nonlinear parameter $\gamma$.
{\em (ii)} Extracted $E_a$ and $\Gamma_0$ as defined by \refE{mainEq}.
{\em (iii)} Verified that the universal scaling of $E_a$ and $\Gamma_0$ initially predicted for a tiny 
region around the bifurcation point holds actually in a region up to two orders 
of magnitude larger than the original one.
{\em (iv)} Verified by solving numerically the exact problem, that the observation is in quantitative agreement with the behavior expected for a driven Duffing oscillator. The scaling of $E_a$ as a function of $|\Omega-\Omega_b|$ is consistent with the predictions of Refs. \cite{DykScaling,kogan2008}.
Due to the generality of the Duffing model, these results are of interest for a wide class of systems. 
Even beyond the fundamental interest in the scaling laws we point out that the device acts as a 
very sensitive amplifier: 
it allows the detection of tiny variations of the resonator frequency. 
Understanding the frequency fluctuations in mechanical resonators is a current challenge 
of the field \cite{KingPRB2012, ZhangArXiv, LeeuwenArXiv}.
Mastering of the bifurcation escape technique by having a reliable theory and experimental verification
of the scaling of the rates is a crucial step towards the study of modifications induced by other 
phenomena.

We gratefully acknowledge discussions with M. Dykman, K. Hasselbach, E. Lhotel and A. Fefferman. 
We thank J. Minet and C. Guttin for help in setting up the experiment.
We acknowledge support from MICROKELVIN, the EU FRP7 grant 228464 
and of the French ANR grant QNM n$^\circ$ 0404 01.

\bibliography{Fullbib}

\begin{thebibliography}{32}%
\makeatletter
\providecommand \@ifxundefined [1]{%
 \@ifx{#1\undefined}
}%
\providecommand \@ifnum [1]{%
 \ifnum #1\expandafter \@firstoftwo
 \else \expandafter \@secondoftwo
 \fi
}%
\providecommand \@ifx [1]{%
 \ifx #1\expandafter \@firstoftwo
 \else \expandafter \@secondoftwo
 \fi
}%
\providecommand \natexlab [1]{#1}%
\providecommand \enquote  [1]{``#1''}%
\providecommand \bibnamefont  [1]{#1}%
\providecommand \bibfnamefont [1]{#1}%
\providecommand \citenamefont [1]{#1}%
\providecommand \href@noop [0]{\@secondoftwo}%
\providecommand \href [0]{\begingroup \@sanitize@url \@href}%
\providecommand \@href[1]{\@@startlink{#1}\@@href}%
\providecommand \@@href[1]{\endgroup#1\@@endlink}%
\providecommand \@sanitize@url [0]{\catcode `\\12\catcode `\$12\catcode
  `\&12\catcode `\#12\catcode `\^12\catcode `\_12\catcode `\%12\relax}%
\providecommand \@@startlink[1]{}%
\providecommand \@@endlink[0]{}%
\providecommand \url  [0]{\begingroup\@sanitize@url \@url }%
\providecommand \@url [1]{\endgroup\@href {#1}{\urlprefix }}%
\providecommand \urlprefix  [0]{URL }%
\providecommand \Eprint [0]{\href }%
\providecommand \doibase [0]{http://dx.doi.org/}%
\providecommand \selectlanguage [0]{\@gobble}%
\providecommand \bibinfo  [0]{\@secondoftwo}%
\providecommand \bibfield  [0]{\@secondoftwo}%
\providecommand \translation [1]{[#1]}%
\providecommand \BibitemOpen [0]{}%
\providecommand \bibitemStop [0]{}%
\providecommand \bibitemNoStop [0]{.\EOS\space}%
\providecommand \EOS [0]{\spacefactor3000\relax}%
\providecommand \BibitemShut  [1]{\csname bibitem#1\endcsname}%
\let\auto@bib@innerbib\@empty
%</preamble>
\bibitem [{\citenamefont {Arrhenius}(1889)}]{arrhenius}%
  \BibitemOpen
  \bibfield  {author} {\bibinfo {author} {\bibfnamefont {S.}~\bibnamefont
  {Arrhenius}},\ }\href@noop {} {\bibfield  {journal} {\bibinfo  {journal} {Z.
  Physik. Chem.}\ }\textbf {\bibinfo {volume} {4}},\ \bibinfo {pages} {226}
  (\bibinfo {year} {1889})}\BibitemShut {NoStop}%
\bibitem [{\citenamefont {Kramers}(1940)}]{kramers_brownian_1940}%
  \BibitemOpen
  \bibfield  {author} {\bibinfo {author} {\bibfnamefont {H.~A.}\ \bibnamefont
  {Kramers}},\ }\href {\doibase 10.1016/S0031-8914(40)90098-2} {\bibfield
  {journal} {\bibinfo  {journal} {Physica}\ }\textbf {\bibinfo {volume} {7}},\
  \bibinfo {pages} {284} (\bibinfo {year} {1940})}\BibitemShut {NoStop}%
\bibitem [{\citenamefont {Novak}\ \emph {et~al.}(1995)\citenamefont {Novak},
  \citenamefont {Sessoli}, \citenamefont {Caneschi},\ and\ \citenamefont
  {Gatteschi}}]{novak_magnetic_1995}%
  \BibitemOpen
  \bibfield  {author} {\bibinfo {author} {\bibfnamefont {M.~A.}\ \bibnamefont
  {Novak}}, \bibinfo {author} {\bibfnamefont {R.}~\bibnamefont {Sessoli}},
  \bibinfo {author} {\bibfnamefont {A.}~\bibnamefont {Caneschi}}, \ and\
  \bibinfo {author} {\bibfnamefont {D.}~\bibnamefont {Gatteschi}},\ }\href
  {\doibase 10.1016/0304-8853(94)00860-4} {\bibfield  {journal} {\bibinfo
  {journal} {J. of Magn. and Magn. Mater.}\ }\textbf {\bibinfo {volume}
  {146}},\ \bibinfo {pages} {211} (\bibinfo {year} {1995})}\BibitemShut
  {NoStop}%
\bibitem [{\citenamefont {Ozbudak}\ \emph {et~al.}(2004)\citenamefont
  {Ozbudak}, \citenamefont {Thattai}, \citenamefont {Lim}, \citenamefont
  {Shraiman},\ and\ \citenamefont {van
  Oudenaarden}}]{ozbudak_multistability_2004}%
  \BibitemOpen
  \bibfield  {author} {\bibinfo {author} {\bibfnamefont {E.~M.}\ \bibnamefont
  {Ozbudak}}, \bibinfo {author} {\bibfnamefont {M.}~\bibnamefont {Thattai}},
  \bibinfo {author} {\bibfnamefont {H.~N.}\ \bibnamefont {Lim}}, \bibinfo
  {author} {\bibfnamefont {B.~I.}\ \bibnamefont {Shraiman}}, \ and\ \bibinfo
  {author} {\bibfnamefont {A.}~\bibnamefont {van Oudenaarden}},\ }\href
  {\doibase 10.1038/nature02298} {\bibfield  {journal} {\bibinfo  {journal}
  {Nature}\ }\textbf {\bibinfo {volume} {427}},\ \bibinfo {pages} {737}
  (\bibinfo {year} {2004})}\BibitemShut {NoStop}%
\bibitem [{\citenamefont {Dethoff}\ \emph {et~al.}(2012)\citenamefont
  {Dethoff}, \citenamefont {Chugh}, \citenamefont {Mustoe},\ and\ \citenamefont
  {Al-Hashimi}}]{RNA_2004}%
  \BibitemOpen
  \bibfield  {author} {\bibinfo {author} {\bibfnamefont {E.~A.}\ \bibnamefont
  {Dethoff}}, \bibinfo {author} {\bibfnamefont {J.}~\bibnamefont {Chugh}},
  \bibinfo {author} {\bibfnamefont {A.~M.}\ \bibnamefont {Mustoe}}, \ and\
  \bibinfo {author} {\bibfnamefont {H.~M.}\ \bibnamefont {Al-Hashimi}},\ }\href
  {\doibase doi:10.1038/nature10885} {\bibfield  {journal} {\bibinfo  {journal}
  {Nature}\ }\textbf {\bibinfo {volume} {482}},\ \bibinfo {pages} {322}
  (\bibinfo {year} {2012})}\BibitemShut {NoStop}%
\bibitem [{\citenamefont {Turlot}\ \emph {et~al.}(1989)\citenamefont {Turlot},
  \citenamefont {Esteve}, \citenamefont {Urbina}, \citenamefont {Martinis},
  \citenamefont {Devoret}, \citenamefont {Linkwitz},\ and\ \citenamefont
  {Grabert}}]{turlot_escape_1989}%
  \BibitemOpen
  \bibfield  {author} {\bibinfo {author} {\bibfnamefont {E.}~\bibnamefont
  {Turlot}}, \bibinfo {author} {\bibfnamefont {D.}~\bibnamefont {Esteve}},
  \bibinfo {author} {\bibfnamefont {C.}~\bibnamefont {Urbina}}, \bibinfo
  {author} {\bibfnamefont {J.~M.}\ \bibnamefont {Martinis}}, \bibinfo {author}
  {\bibfnamefont {M.~H.}\ \bibnamefont {Devoret}}, \bibinfo {author}
  {\bibfnamefont {S.}~\bibnamefont {Linkwitz}}, \ and\ \bibinfo {author}
  {\bibfnamefont {H.}~\bibnamefont {Grabert}},\ }\href {\doibase
  10.1103/PhysRevLett.62.1788} {\bibfield  {journal} {\bibinfo  {journal}
  {Phys. Rev. Lett.}\ }\textbf {\bibinfo {volume} {62}},\ \bibinfo {pages}
  {1788} (\bibinfo {year} {1989})}\BibitemShut {NoStop}%
\bibitem [{\citenamefont {Kurkijärvi}(1972)}]{kurkijarvi_intrinsic_1972}%
  \BibitemOpen
  \bibfield  {author} {\bibinfo {author} {\bibfnamefont {J.}~\bibnamefont
  {Kurkijärvi}},\ }\href {\doibase 10.1103/PhysRevB.6.832} {\bibfield
  {journal} {\bibinfo  {journal} {Phys. Rev. B}\ }\textbf {\bibinfo {volume}
  {6}},\ \bibinfo {pages} {832} (\bibinfo {year} {1972})}\BibitemShut {NoStop}%
\bibitem [{\citenamefont {Lapidus}\ \emph {et~al.}(1999)\citenamefont
  {Lapidus}, \citenamefont {Enzer},\ and\ \citenamefont
  {Gabrielse}}]{lapidus_stochastic_1999}%
  \BibitemOpen
  \bibfield  {author} {\bibinfo {author} {\bibfnamefont {L.~J.}\ \bibnamefont
  {Lapidus}}, \bibinfo {author} {\bibfnamefont {D.}~\bibnamefont {Enzer}}, \
  and\ \bibinfo {author} {\bibfnamefont {G.}~\bibnamefont {Gabrielse}},\ }\href
  {\doibase 10.1103/PhysRevLett.83.899} {\bibfield  {journal} {\bibinfo
  {journal} {Phys. Rev. Lett.}\ }\textbf {\bibinfo {volume} {83}},\ \bibinfo
  {pages} {899} (\bibinfo {year} {1999})}\BibitemShut {NoStop}%
\bibitem [{\citenamefont {Siddiqi}\ \emph {et~al.}(2004)\citenamefont
  {Siddiqi}, \citenamefont {Vijay}, \citenamefont {Pierre}, \citenamefont
  {Wilson}, \citenamefont {Metcalfe}, \citenamefont {Rigetti}, \citenamefont
  {Frunzio},\ and\ \citenamefont {Devoret}}]{siddiqi_rf-driven_2004}%
  \BibitemOpen
  \bibfield  {author} {\bibinfo {author} {\bibfnamefont {I.}~\bibnamefont
  {Siddiqi}}, \bibinfo {author} {\bibfnamefont {R.}~\bibnamefont {Vijay}},
  \bibinfo {author} {\bibfnamefont {F.}~\bibnamefont {Pierre}}, \bibinfo
  {author} {\bibfnamefont {C.~M.}\ \bibnamefont {Wilson}}, \bibinfo {author}
  {\bibfnamefont {M.}~\bibnamefont {Metcalfe}}, \bibinfo {author}
  {\bibfnamefont {C.}~\bibnamefont {Rigetti}}, \bibinfo {author} {\bibfnamefont
  {L.}~\bibnamefont {Frunzio}}, \ and\ \bibinfo {author} {\bibfnamefont
  {M.~H.}\ \bibnamefont {Devoret}},\ }\href {\doibase
  10.1103/PhysRevLett.93.207002} {\bibfield  {journal} {\bibinfo  {journal}
  {Phys. Rev. Lett.}\ }\textbf {\bibinfo {volume} {93}},\ \bibinfo {pages}
  {207002} (\bibinfo {year} {2004})}\BibitemShut {NoStop}%
\bibitem [{\citenamefont {Aldridge}\ and\ \citenamefont
  {Cleland}(2005)}]{aldridge_noise-enabled_2005}%
  \BibitemOpen
  \bibfield  {author} {\bibinfo {author} {\bibfnamefont {J.~S.}\ \bibnamefont
  {Aldridge}}\ and\ \bibinfo {author} {\bibfnamefont {A.~N.}\ \bibnamefont
  {Cleland}},\ }\href {\doibase 10.1103/PhysRevLett.94.156403} {\bibfield
  {journal} {\bibinfo  {journal} {Phys. Rev. Lett.}\ }\textbf {\bibinfo
  {volume} {94}},\ \bibinfo {pages} {156403} (\bibinfo {year}
  {2005})}\BibitemShut {NoStop}%
\bibitem [{\citenamefont {Stambaugh}\ and\ \citenamefont
  {Chan}(2006)}]{stambaugh_noise-activated_2006}%
  \BibitemOpen
  \bibfield  {author} {\bibinfo {author} {\bibfnamefont {C.}~\bibnamefont
  {Stambaugh}}\ and\ \bibinfo {author} {\bibfnamefont {H.~B.}\ \bibnamefont
  {Chan}},\ }\href {\doibase 10.1103/PhysRevB.73.172302} {\bibfield  {journal}
  {\bibinfo  {journal} {Phys. Rev. B}\ }\textbf {\bibinfo {volume} {73}},\
  \bibinfo {pages} {172302} (\bibinfo {year} {2006})}\BibitemShut {NoStop}%
\bibitem [{\citenamefont {Chan}\ and\ \citenamefont
  {Stambaugh}(2007)}]{chan_activation_2007}%
  \BibitemOpen
  \bibfield  {author} {\bibinfo {author} {\bibfnamefont {H.~B.}\ \bibnamefont
  {Chan}}\ and\ \bibinfo {author} {\bibfnamefont {C.}~\bibnamefont
  {Stambaugh}},\ }\href {\doibase 10.1103/PhysRevLett.99.060601} {\bibfield
  {journal} {\bibinfo  {journal} {Phys. Rev. Lett.}\ }\textbf {\bibinfo
  {volume} {99}},\ \bibinfo {pages} {060601} (\bibinfo {year}
  {2007})}\BibitemShut {NoStop}%
\bibitem [{\citenamefont {Chan}\ \emph {et~al.}(2008)\citenamefont {Chan},
  \citenamefont {Dykman},\ and\ \citenamefont {Stambaugh}}]{chan_paths_2008}%
  \BibitemOpen
  \bibfield  {author} {\bibinfo {author} {\bibfnamefont {H.~B.}\ \bibnamefont
  {Chan}}, \bibinfo {author} {\bibfnamefont {M.~I.}\ \bibnamefont {Dykman}}, \
  and\ \bibinfo {author} {\bibfnamefont {C.}~\bibnamefont {Stambaugh}},\ }\href
  {\doibase 10.1103/PhysRevLett.100.130602} {\bibfield  {journal} {\bibinfo
  {journal} {Phys. Rev. Lett.}\ }\textbf {\bibinfo {volume} {100}},\ \bibinfo
  {pages} {130602} (\bibinfo {year} {2008})}\BibitemShut {NoStop}%
\bibitem [{\citenamefont {Unterreithmeier}\ \emph {et~al.}(2010)\citenamefont
  {Unterreithmeier}, \citenamefont {Faust},\ and\ \citenamefont
  {Kotthaus}}]{unterreithmeier_nonlinear_2010}%
  \BibitemOpen
  \bibfield  {author} {\bibinfo {author} {\bibfnamefont {Q.~P.}\ \bibnamefont
  {Unterreithmeier}}, \bibinfo {author} {\bibfnamefont {T.}~\bibnamefont
  {Faust}}, \ and\ \bibinfo {author} {\bibfnamefont {J.~P.}\ \bibnamefont
  {Kotthaus}},\ }\href {\doibase 10.1103/PhysRevB.81.241405} {\bibfield
  {journal} {\bibinfo  {journal} {Phys. Rev. B}\ }\textbf {\bibinfo {volume}
  {81}},\ \bibinfo {pages} {241405} (\bibinfo {year} {2010})}\BibitemShut
  {NoStop}%
\bibitem [{\citenamefont {Boulant}\ \emph {et~al.}(2007)\citenamefont
  {Boulant}, \citenamefont {Ithier}, \citenamefont {Meeson}, \citenamefont
  {Nguyen}, \citenamefont {Vion}, \citenamefont {Esteve}, \citenamefont
  {Siddiqi}, \citenamefont {Vijay}, \citenamefont {Rigetti}, \citenamefont
  {Pierre},\ and\ \citenamefont {Devoret}}]{boulant_quantum_2007}%
  \BibitemOpen
  \bibfield  {author} {\bibinfo {author} {\bibfnamefont {N.}~\bibnamefont
  {Boulant}}, \bibinfo {author} {\bibfnamefont {G.}~\bibnamefont {Ithier}},
  \bibinfo {author} {\bibfnamefont {P.}~\bibnamefont {Meeson}}, \bibinfo
  {author} {\bibfnamefont {F.}~\bibnamefont {Nguyen}}, \bibinfo {author}
  {\bibfnamefont {D.}~\bibnamefont {Vion}}, \bibinfo {author} {\bibfnamefont
  {D.}~\bibnamefont {Esteve}}, \bibinfo {author} {\bibfnamefont
  {I.}~\bibnamefont {Siddiqi}}, \bibinfo {author} {\bibfnamefont
  {R.}~\bibnamefont {Vijay}}, \bibinfo {author} {\bibfnamefont
  {C.}~\bibnamefont {Rigetti}}, \bibinfo {author} {\bibfnamefont
  {F.}~\bibnamefont {Pierre}}, \ and\ \bibinfo {author} {\bibfnamefont
  {M.}~\bibnamefont {Devoret}},\ }\href {\doibase 10.1103/PhysRevB.76.014525}
  {\bibfield  {journal} {\bibinfo  {journal} {Phys. Rev. B}\ }\textbf {\bibinfo
  {volume} {76}},\ \bibinfo {pages} {014525} (\bibinfo {year}
  {2007})}\BibitemShut {NoStop}%
\bibitem [{\citenamefont {H\"anggi}\ \emph {et~al.}(1990)\citenamefont
  {H\"anggi}, \citenamefont {Talkner},\ and\ \citenamefont
  {Borkovec}}]{hanggi_reaction-rate_1990}%
  \BibitemOpen
  \bibfield  {author} {\bibinfo {author} {\bibfnamefont {P.}~\bibnamefont
  {H\"anggi}}, \bibinfo {author} {\bibfnamefont {P.}~\bibnamefont {Talkner}}, \
  and\ \bibinfo {author} {\bibfnamefont {M.}~\bibnamefont {Borkovec}},\ }\href
  {\doibase 10.1103/RevModPhys.62.251} {\bibfield  {journal} {\bibinfo
  {journal} {Rev. Mod. Phys.}\ }\textbf {\bibinfo {volume} {62}},\ \bibinfo
  {pages} {251} (\bibinfo {year} {1990})}\BibitemShut {NoStop}%
\bibitem [{\citenamefont {Dykman}\ and\ \citenamefont
  {Krivoglaz}(1979)}]{dykman_theory_1979}%
  \BibitemOpen
  \bibfield  {author} {\bibinfo {author} {\bibfnamefont {M.~I.}\ \bibnamefont
  {Dykman}}\ and\ \bibinfo {author} {\bibfnamefont {M.~A.}\ \bibnamefont
  {Krivoglaz}},\ }\href@noop {} {\bibfield  {journal} {\bibinfo  {journal}
  {Sov. Phys. {JETP}}\ ,\ \bibinfo {pages} {60}} (\bibinfo {year}
  {1979})}\BibitemShut {NoStop}%
\bibitem [{\citenamefont {Dykman}\ \emph {et~al.}(2005)\citenamefont {Dykman},
  \citenamefont {Schwartz},\ and\ \citenamefont {Shapiro}}]{DykScaling}%
  \BibitemOpen
  \bibfield  {author} {\bibinfo {author} {\bibfnamefont {M.}~\bibnamefont
  {Dykman}}, \bibinfo {author} {\bibfnamefont {I.}~\bibnamefont {Schwartz}}, \
  and\ \bibinfo {author} {\bibfnamefont {M.}~\bibnamefont {Shapiro}},\
  }\href@noop {} {\bibfield  {journal} {\bibinfo  {journal} {Phys. Rev. E}\
  }\textbf {\bibinfo {volume} {72}},\ \bibinfo {pages} {021102} (\bibinfo
  {year} {2005})}\BibitemShut {NoStop}%
\bibitem [{\citenamefont {Kogan}(2008)}]{kogan2008}%
  \BibitemOpen
  \bibfield  {author} {\bibinfo {author} {\bibfnamefont {O.}~\bibnamefont
  {Kogan}},\ }\href@noop {} {\bibfield  {journal} {\bibinfo  {journal}
  {arXiv:0805.0972v2}\ } (\bibinfo {year} {2008})}\BibitemShut {NoStop}%
\bibitem [{\citenamefont {Collin}\ \emph {et~al.}(2012)\citenamefont {Collin},
  \citenamefont {Defoort}, \citenamefont {Lulla}, \citenamefont {Moutonet},
  \citenamefont {Heron}, \citenamefont {Bourgeois}, \citenamefont {Bunkov},\
  and\ \citenamefont {Godfrin}}]{collin_-situ_2012}%
  \BibitemOpen
  \bibfield  {author} {\bibinfo {author} {\bibfnamefont {E.}~\bibnamefont
  {Collin}}, \bibinfo {author} {\bibfnamefont {M.}~\bibnamefont {Defoort}},
  \bibinfo {author} {\bibfnamefont {K.}~\bibnamefont {Lulla}}, \bibinfo
  {author} {\bibfnamefont {T.}~\bibnamefont {Moutonet}}, \bibinfo {author}
  {\bibfnamefont {J.-S.}\ \bibnamefont {Heron}}, \bibinfo {author}
  {\bibfnamefont {O.}~\bibnamefont {Bourgeois}}, \bibinfo {author}
  {\bibfnamefont {Y.~M.}\ \bibnamefont {Bunkov}}, \ and\ \bibinfo {author}
  {\bibfnamefont {H.}~\bibnamefont {Godfrin}},\ }\href {\doibase
  10.1063/1.4705992} {\bibfield  {journal} {\bibinfo  {journal} {Review of
  Scientific Instruments}\ }\textbf {\bibinfo {volume} {83}},\ \bibinfo {pages}
  {045005} (\bibinfo {year} {2012})}\BibitemShut {NoStop}%
\bibitem [{\citenamefont {Cleland}\ and\ \citenamefont
  {Roukes}(1999)}]{clelandmagneto}%
  \BibitemOpen
  \bibfield  {author} {\bibinfo {author} {\bibfnamefont {A.}~\bibnamefont
  {Cleland}}\ and\ \bibinfo {author} {\bibfnamefont {M.}~\bibnamefont
  {Roukes}},\ }\href@noop {} {\bibfield  {journal} {\bibinfo  {journal}
  {Sensors and Actuators}\ }\textbf {\bibinfo {volume} {72}},\ \bibinfo {pages}
  {256} (\bibinfo {year} {1999})}\BibitemShut {NoStop}%
\bibitem [{\citenamefont {Collin}\ \emph
  {et~al.}(2011{\natexlab{a}})\citenamefont {Collin}, \citenamefont {Moutonet},
  \citenamefont {Heron}, \citenamefont {Bourgeois}, \citenamefont {Bunkov},\
  and\ \citenamefont {Godfrin}}]{collin_tunable_2011}%
  \BibitemOpen
  \bibfield  {author} {\bibinfo {author} {\bibfnamefont {E.}~\bibnamefont
  {Collin}}, \bibinfo {author} {\bibfnamefont {T.}~\bibnamefont {Moutonet}},
  \bibinfo {author} {\bibfnamefont {J.-S.}\ \bibnamefont {Heron}}, \bibinfo
  {author} {\bibfnamefont {O.}~\bibnamefont {Bourgeois}}, \bibinfo {author}
  {\bibfnamefont {Y.~M.}\ \bibnamefont {Bunkov}}, \ and\ \bibinfo {author}
  {\bibfnamefont {H.}~\bibnamefont {Godfrin}},\ }\href {\doibase
  10.1007/s10909-010-0257-5} {\bibfield  {journal} {\bibinfo  {journal} {J Low
  Temp Phys}\ }\textbf {\bibinfo {volume} {162}},\ \bibinfo {pages} {653}
  (\bibinfo {year} {2011}{\natexlab{a}})}\BibitemShut {NoStop}%
\bibitem [{\citenamefont {Kozinsky}\ \emph {et~al.}(2006)\citenamefont
  {Kozinsky}, \citenamefont {Postma}, \citenamefont {Bargatin},\ and\
  \citenamefont {Roukes}}]{kozinsky_tuning_2006}%
  \BibitemOpen
  \bibfield  {author} {\bibinfo {author} {\bibfnamefont {I.}~\bibnamefont
  {Kozinsky}}, \bibinfo {author} {\bibfnamefont {H.~W.~C.}\ \bibnamefont
  {Postma}}, \bibinfo {author} {\bibfnamefont {I.}~\bibnamefont {Bargatin}}, \
  and\ \bibinfo {author} {\bibfnamefont {M.~L.}\ \bibnamefont {Roukes}},\
  }\href {\doibase 10.1063/1.2209211} {\bibfield  {journal} {\bibinfo
  {journal} {Applied Physics Letters}\ }\textbf {\bibinfo {volume} {88}},\
  \bibinfo {pages} {253101} (\bibinfo {year} {2006})}\BibitemShut {NoStop}%
\bibitem [{\citenamefont {Zaitsev}\ \emph {et~al.}(2012)\citenamefont
  {Zaitsev}, \citenamefont {Shtempluck}, \citenamefont {Buks},\ and\
  \citenamefont {Gottlieb}}]{NonlinDamping}%
  \BibitemOpen
  \bibfield  {author} {\bibinfo {author} {\bibfnamefont {S.}~\bibnamefont
  {Zaitsev}}, \bibinfo {author} {\bibfnamefont {O.}~\bibnamefont {Shtempluck}},
  \bibinfo {author} {\bibfnamefont {E.}~\bibnamefont {Buks}}, \ and\ \bibinfo
  {author} {\bibfnamefont {O.}~\bibnamefont {Gottlieb}},\ }\href@noop {}
  {\bibfield  {journal} {\bibinfo  {journal} {Nonlinear Dynamics}\ }\textbf
  {\bibinfo {volume} {67}},\ \bibinfo {pages} {859} (\bibinfo {year}
  {2012})}\BibitemShut {NoStop}%
\bibitem [{\citenamefont {Collin}\ \emph
  {et~al.}(2011{\natexlab{b}})\citenamefont {Collin}, \citenamefont {Moutonet},
  \citenamefont {Heron}, \citenamefont {Bourgeois}, \citenamefont {Bunkov},\
  and\ \citenamefont {Godfrin}}]{parametric}%
  \BibitemOpen
  \bibfield  {author} {\bibinfo {author} {\bibfnamefont {E.}~\bibnamefont
  {Collin}}, \bibinfo {author} {\bibfnamefont {T.}~\bibnamefont {Moutonet}},
  \bibinfo {author} {\bibfnamefont {J.-S.}\ \bibnamefont {Heron}}, \bibinfo
  {author} {\bibfnamefont {O.}~\bibnamefont {Bourgeois}}, \bibinfo {author}
  {\bibfnamefont {Y.~M.}\ \bibnamefont {Bunkov}}, \ and\ \bibinfo {author}
  {\bibfnamefont {H.}~\bibnamefont {Godfrin}},\ }\href@noop {} {\bibfield
  {journal} {\bibinfo  {journal} {Phys. Rev. B}\ }\textbf {\bibinfo {volume}
  {84}},\ \bibinfo {pages} {054108} (\bibinfo {year}
  {2011}{\natexlab{b}})}\BibitemShut {NoStop}%
\bibitem [{\citenamefont {Collin}\ \emph {et~al.}(2010)\citenamefont {Collin},
  \citenamefont {Bunkov},\ and\ \citenamefont
  {Godfrin}}]{collin_addressing_2010}%
  \BibitemOpen
  \bibfield  {author} {\bibinfo {author} {\bibfnamefont {E.}~\bibnamefont
  {Collin}}, \bibinfo {author} {\bibfnamefont {Y.~M.}\ \bibnamefont {Bunkov}},
  \ and\ \bibinfo {author} {\bibfnamefont {H.}~\bibnamefont {Godfrin}},\ }\href
  {\doibase 10.1103/PhysRevB.82.235416} {\bibfield  {journal} {\bibinfo
  {journal} {Phys. Rev. B}\ }\textbf {\bibinfo {volume} {82}},\ \bibinfo
  {pages} {235416} (\bibinfo {year} {2010})}\BibitemShut {NoStop}%
\bibitem [{\citenamefont {Fong}\ \emph {et~al.}(2012)\citenamefont {Fong},
  \citenamefont {Pernice},\ and\ \citenamefont {Tang}}]{KingPRB2012}%
  \BibitemOpen
  \bibfield  {author} {\bibinfo {author} {\bibfnamefont {K.~Y.}\ \bibnamefont
  {Fong}}, \bibinfo {author} {\bibfnamefont {W.~H.~P.}\ \bibnamefont
  {Pernice}}, \ and\ \bibinfo {author} {\bibfnamefont {H.~X.}\ \bibnamefont
  {Tang}},\ }\href {\doibase 10.1103/PhysRevB.85.161410} {\bibfield  {journal}
  {\bibinfo  {journal} {Phys. Rev. B}\ }\textbf {\bibinfo {volume} {85}},\
  \bibinfo {pages} {161410(R)} (\bibinfo {year} {2012})}\BibitemShut {NoStop}%
\bibitem [{\citenamefont {Zhang}\ \emph {et~al.}(2014)\citenamefont {Zhang},
  \citenamefont {Moser}, \citenamefont {G\"uttinger}, \citenamefont
  {Bachtold},\ and\ \citenamefont {Dykman}}]{ZhangArXiv}%
  \BibitemOpen
  \bibfield  {author} {\bibinfo {author} {\bibfnamefont {Y.}~\bibnamefont
  {Zhang}}, \bibinfo {author} {\bibfnamefont {J.}~\bibnamefont {Moser}},
  \bibinfo {author} {\bibfnamefont {J.}~\bibnamefont {G\"uttinger}}, \bibinfo
  {author} {\bibfnamefont {A.}~\bibnamefont {Bachtold}}, \ and\ \bibinfo
  {author} {\bibfnamefont {M.~I.}\ \bibnamefont {Dykman}},\ }\href@noop {}
  {\bibfield  {journal} {\bibinfo  {journal} {Phys. Rev. Lett.}\ }\textbf
  {\bibinfo {volume} {113}},\ \bibinfo {pages} {255502} (\bibinfo {year}
  {2014})}\BibitemShut {NoStop}%
\bibitem [{\citenamefont {van Leeuwen}\ \emph {et~al.}(2014)\citenamefont {van
  Leeuwen}, \citenamefont {Castellanos-Gomez}, \citenamefont {Steele},
  \citenamefont {van~der Zant},\ and\ \citenamefont {Venstra}}]{LeeuwenArXiv}%
  \BibitemOpen
  \bibfield  {author} {\bibinfo {author} {\bibfnamefont {R.}~\bibnamefont {van
  Leeuwen}}, \bibinfo {author} {\bibfnamefont {A.}~\bibnamefont
  {Castellanos-Gomez}}, \bibinfo {author} {\bibfnamefont {G.}~\bibnamefont
  {Steele}}, \bibinfo {author} {\bibfnamefont {H.}~\bibnamefont {van~der
  Zant}}, \ and\ \bibinfo {author} {\bibfnamefont {W.}~\bibnamefont
  {Venstra}},\ }\href@noop {} {\bibfield  {journal} {\bibinfo  {journal} {Appl.
  Phys. Lett.}\ }\textbf {\bibinfo {volume} {105}},\ \bibinfo {pages} {041911}
  (\bibinfo {year} {2014})}\BibitemShut {NoStop}%
\bibitem [{\citenamefont {Eadie}\ \emph {et~al.}(1971)\citenamefont {Eadie},
  \citenamefont {Drijard}, \citenamefont {James}, \citenamefont {Roos},\ and\
  \citenamefont {Sadoulet}}]{KolmoSmirn}%
  \BibitemOpen
  \bibfield  {author} {\bibinfo {author} {\bibfnamefont {W.}~\bibnamefont
  {Eadie}}, \bibinfo {author} {\bibfnamefont {D.}~\bibnamefont {Drijard}},
  \bibinfo {author} {\bibfnamefont {F.}~\bibnamefont {James}}, \bibinfo
  {author} {\bibfnamefont {M.}~\bibnamefont {Roos}}, \ and\ \bibinfo {author}
  {\bibfnamefont {B.}~\bibnamefont {Sadoulet}},\ }\href@noop {} {\emph
  {\bibinfo {title} {Statistical Methods in Experimental Physics}}},\ edited
  by\ \bibinfo {editor} {\bibnamefont {North-Holland}}\ (\bibinfo {address}
  {Amsterdam},\ \bibinfo {year} {1971})\BibitemShut {NoStop}%
\bibitem [{\citenamefont {Defoort}(2014)}]{MartialPHD}%
  \BibitemOpen
  \bibfield  {author} {\bibinfo {author} {\bibfnamefont {M.}~\bibnamefont
  {Defoort}},\ }\href@noop {} {\emph {\bibinfo {title} {PhD thesis: Non-linear
  dynamics in nano-electromechanical systems at low temperature}}}\ (\bibinfo
  {address} {CNRS et Universit\'e Grenoble Alpes, unpublished},\ \bibinfo
  {year} {2014})\BibitemShut {NoStop}%
\bibitem [{\citenamefont {Pistolesi}\ \emph {et~al.}(2008)\citenamefont
  {Pistolesi}, \citenamefont {Blanter},\ and\ \citenamefont
  {Martin}}]{pistolesi_self-consistent_2008}%
  \BibitemOpen
  \bibfield  {author} {\bibinfo {author} {\bibfnamefont {F.}~\bibnamefont
  {Pistolesi}}, \bibinfo {author} {\bibfnamefont {Y.}~\bibnamefont {Blanter}},
  \ and\ \bibinfo {author} {\bibfnamefont {I.}~\bibnamefont {Martin}},\ }\href
  {http://link.aps.org/doi/10.1103/PhysRevB.78.085127} {\bibfield  {journal}
  {\bibinfo  {journal} {Physical Review B}\ }\textbf {\bibinfo {volume} {78}},\
  \bibinfo {pages} {085127} (\bibinfo {year} {2008})}\BibitemShut {NoStop}%
\end{thebibliography}%

\end{document}